\documentclass[aps,pra,showpacs,twocolumn,amsmath,preprintnumbers,nofootinbib,secnumarabic,a4paper,floatfix]{revtex4} 

\usepackage{epsfig,verbatim}
\usepackage{amssymb,amsfonts}
\usepackage{hyperref}

\usepackage[caption=false]{subfig}
\def\sc{0.49}  
\def\lsc{0.488}\def\rsc{0.29} 

\def\art{paper}

\def\jrn#1#2#3#4#5#6{#1 \textit{#3} \textbf{#4} #5 (#6)} \def\boo#1#2#3#4#5#6#7{#1 \textit{#2} (#5: #3) #7 (#6)}  \def\andd{ and } \def\andt{ and } \def\eq{Eq. }   \def\Ref{Ref. }  \def\Fig{Fig. } 

\def\eqref#1{(\ref{#1})}
\def\scn#1#2{\section{#1}\lb{#2}}
\def\sscn#1#2{\subsection{#1}\lb{#2}}
\def\bfl{\begin{flushleft}}
\def\efl{\end{flushleft}}
\def\bfr{\begin{flushright}}
\def\efr{\end{flushright}}
\def\bc{\begin{center}}
\def\ec{\end{center}}
\def\be{\begin{equation}}
\def\ee{\end{equation}}
\def\bse{\begin{subequations}}
\def\ese{\end{subequations}}
\def\ba{\begin{eqnarray}}
\def\ea{\end{eqnarray}}
\def\baa#1{\begin{array}{#1}}
\def\eaa{\end{array}}
\def\bw{\begin{widetext}}
\def\ew{\end{widetext}}

\def\lb#1{\label{#1}}
\def\bit{\begin{itemize}}
\def\eit{\end{itemize}}
\def\bco{}
\def\bcs{\begin{cases}}
\def\ecs{\end{cases}}

\def\vena{\boldsymbol{\nabla}}

\def\cf{{\cal D}}
\def\vol{{\cal V}}
\def\vol{V}
\def\dvol{\text{d}\vol}
\def\lanp{{\cal V}}
\def\boldsymbol#1{#1}
\def\dm{{\bar d}}

\begin{document}

\preprint{\small Indian J. Phys. \textbf{96}, 2385-2392 (2022)   
\quad 
[\href{https://doi.org/10.1007/s12648-021-02190-2}{DOI: 10.1007/s12648-021-02190-2}]
}

\title{
Logarithmic wave-mechanical effects in polycrystalline metals: Theory and experiment
}

\author{Maksym Kraiev}
\affiliation{Yuzhnoye State Design Office, 
Dnipro 49008, Ukraine}

\author{Kateryna Domina}
\affiliation{Z. I. Nekrasov Iron and Steel Institute of NAS of Ukraine, 
Dnipro 49107, Ukraine}

\author{Violeta Kraieva}
\affiliation{Department of Physics, 
Dnipro National University of Railway Transport, 
Dnipro 49010, Ukraine}

\author{Konstantin G. Zloshchastiev}
\email{https://bit.do/kgz}
\affiliation{Institute of Systems Science, Durban University of Technology, 
Durban 4000, South Africa}



\begin{abstract} 
Schr\"odinger-type wave equations with logarithmic nonlinearity occur in hydrodynamic models of
Korteweg-type materials with capillarity and surface tension, 
which can undergo liquid-solid or liquid-gas phase transitions. 
One of the predictions of the theory is a periodic pattern of density inhomogeneities 
occurring in the form of either bubbles (topological phase), or cells (non-topological phase).
Such inhomogeneities are described
by solitonic solutions 
of a logarithmic wave equation, gaussons and kinks,
in the vicinity of the liquid-solid phase transition. 
During the solidification process, these inhomogeneities become centers of nucleation,
thus shaping the polycrystalline structure of the metal grains.
The theory predicts a Gaussian profile of material density inside such a cell,
which should manifest in a Gaussian-like profile of microhardness inside a grain.
We report experimental evidence of large-scale periodicity in 
the structure of grains in the ferrite steel S235/A570, copper C-Cu/C14200, austenite in steel X10CrNiTi18-10/AISI 321, and aluminium-magnesium alloy 5083/5056;
and also Gaussian-like profiles of microhardness inside an averaged grain in these materials.
\end{abstract}

\date{received: 9 November 2020 [Springer]}


\maketitle

\scn{Introduction}{s:intr}
Various crystalline structures occur in metals during 
the process of solidification.
In pure metals, this is a simple crystal lattice, which can be
of a cubic, body-centered cubic or face-centered cubic type.
In real metals, 
a more complicated picture occurs, due to the presence of impurities, defects and other factors.
Throughout the melt's bulk, multiple centers of nucleation occur,
so that
crystals begin to grow from those centers outwards,
until their boundaries reach each other.
This is where interfaces develop between the borders of single-crystal domain grains.
Finally, a polycrystalline structure occurs as a large-scale
pattern of adjacent grains \cite{sal64,nz66,fle74}. 

This leads to our first questions: how orderly is this pattern,
and,
is the size and mass of each grain a
completely random value? 
Aside from purely theoretical interest, such questions have immediate practical importance.
The microhardness of grains, 
their size and density distributions, are very important properties of polycrystalline metals and alloys,
because they strongly affect the macroscopic properties thereof, such as average hardness, plasticity, viscosity and 
mass \cite{gc19}.

These properties are formed when a grain nucleates from the molt during the solidification process.
Using  conventional statistical mechanics for such systems would result in a very cumbersome description,
due to the complexity 
of its physical and chemical components.
One could try to apply a ``common wisdom'' reasoning: 
because numerous impurities are randomly distributed in the molten bulk,
and so are the associated nucleation centers,
distances between those centers are expected to be random too.
Consequently, one might anticipate that the distribution of grains' sizes would be uniform; or at least, would
have some wide plateau-like distribution,
so that no periodic large-scale pattern 
can form in the polycrystalline metals. 
But is this what actually happens?

It turns out that it is not.
Numerous experimental data and metallographic analysis reveal that
the distribution of grain sizes in cast metals is predominantly Gaussian; although
it also depends on how the material is treated.
Thus, during the collective recrystallization process, the distribution usually has a Gaussian form,
with a shift towards larger sizes as temperature and soaking time on heating increase \cite{sal58,gor78,lb04}.
However, this distribution and relevant formulae have been established in those materials phenomenologically,
while a proper theoretical explanation of the Gaussian behaviour of grain size distributions is 
still pending.

This raises a further question:
is there an analytical way of explaining the matters raised above?
For example, is it possible to find a set of effective collective degrees of freedom,
in which the description of the materials in question can be substantially simplified,
other
than by using \textit{ab initio} many-body statistical mechanics?
Is it possible to find any universal laws and relations pertinent to the systems in question,
or, at least, to a wide range thereof?

In this \art,
we attempt to answer these questions, both on theoretical grounds, and
by doing
experimental studies
of
non-alloy structural steel S235/A570, copper C-Cu/C14200, stainless steel X10CrNiTi18-10/AISI 321, and aluminium-magnesium alloy 5083/5056.
The paper is organized as follows.
In Sec. \ref{s:mod}, we give an overview of the logarithmic Korteweg-type model,
both in its wave-mechanical and hydrodynamic representations,
we also describe its phase structure. 
The predictions of the model pertinent to polycrystalline metallic materials
are discussed in Sec. \ref{s:thp}.
The experimental setup and data are described in Sec. \ref{s:ex},
where we used the Vickers hardness method
for microhardness measurements of microstructure components.
Discussion of results and conclusions are presented in Sec. \ref{s:dis}.

\scn{The model}{s:mod}
To address the problems mentioned in the Introduction,
we use the wave-mechanical approach 
based on wave equations with logarithmic nonlinearity.
This 
approach finds fruitful applications in the physics of condensed matter,
to mention just a few of the literature
landmarks 
\cite{dmf03,z11ap,az11,z12eb,z17zna,z18zna,sz19,z19ijmpb}.
The wide applicability and universality of logarithmic models can be explained by the fact 
that logarithmic nonlinearity 
appears in a leading-order approximation in the effective description
of those strongly-interacting many-body systems
in which interaction energies predominate kinetic ones,
and which allow a fluid-mechanical description,
see \Ref\cite{z18zna} for details.
Examples of such systems include 
not only low-temperature quantum Bose liquids, such as dense Bose-Einstein condensates \cite{az11,z17zna} or superfluid helium-4 \cite{z12eb,sz19,z19ijmpb};
but also 
Korteweg-type materials which can undergo liquid-solid or liquid-gas phase transitions \cite{dmf03,z18epl}. 
In these materials, capillarity and surface tension play a substantial role, which makes them
useful for modeling various flows 
with non-negligible surface and interface effects \cite{ds85,amw98}.

One of the theory's predictions is inhomogeneities of 
density, 
caused by the existence of
multiple soliton and Gaussian-shaped solitary wave solutions for an underlying logarithmic wave equation
in the vicinity of a liquid-solid phase transition. 
Each of these inhomogeneities
can manifest themselves
in the form of ``bubbles'' (density is smaller at the center of a domain than at its boundaries)
or ``cells'' (density is larger at the center than at the boundaries).
Recent studies of these phenomena were done for natural silicate materials 
in geophysics \cite{z18epl}, such as magmas
in volcanic conduits, where 
it is well-known that (approximately) periodical flows and structures occur \cite{dmf03}.

\sscn{Wave-mechanical representation}{s:wmf}
When molten metals cool to temperatures near their solidification point,
the
characteristic kinetic energies of their atoms become smaller than the interaction potentials between them.
This can happen, not only at low temperatures, but also in high-density or
effectively low-dimensional systems.
Moreover, 
such materials behave more like fluids
than solids or gases.

Therefore,
following the line of reasoning of works \cite{z18zna,z18epl},
we assume that,
in a leading-order approximation:
(i) one can apply a hydrodynamical description to such materials,
by introducing a fluid function, which encodes the main 
properties of a flow, such as density and velocity,
(ii) the fluid function's dynamics is governed by a wave-mechanical equation,
which somewhat resembles the Schr\"odinger equation, but with a nonlinear term, 
and (iii) logarithmic nonlinearity occurs in this equation.

This equation can thus be written in the form 
\ba
i \partial_t \Psi
=
\left[-\frac{\cf}{2} \vena^2
- b 
 \ln\left(|\Psi|^{2}/\rho_0\right)
\right]\Psi
,\label{e:o}
\ea
where 
$b$, $\rho_0$ and $\cf$ are real-valued parameters.
Here, the fluid wave-function
$\Psi (\textbf{x},t)$
contains information about the fluid's macroscopic parameters,
such as density and flow velocity;
general discussion of such functions can be found in \Ref \cite{ry99}, for example.
One can show that the nonlinear coupling $b$ is a linear function of temperature,
$b  \sim T$, 
therefore, by setting $b = \text{const}$ from now on we assume that our fluid is in an isothermal state \cite{z18zna}.

Equation \eqref{e:o} 
must be supplemented with the normalization condition
\be\lb{e:norm}
\int_\vol |\Psi|^2 \dvol  = 
\int_\vol \rho\, \dvol = M
,
\ee 
where $\rho = \rho(\textbf{x},t)$ is fluid density,
$M$ and $\vol$ are the total mass and volume of the fluid.
This condition leads to analytical restrictions upon fluid functions:
for the same set of 
boundary conditions, \eq \eqref{e:o}
allows multiple normalizable (eigen)solutions
which
correspond to excited states in the associated Hilbert space of the problem, for example $L^2$.
In other words,
a set of all 
fluid wavefunctions must
form a Hilbert space,
which can be interpreted as this fluid spontaneously ``choosing'' to be in one of these states 
(the ground state is still preferred, because it corresponds to the
minimum of wave-mechanical energy). 

Thus,
the fluid function description is somewhat similar 
to the formalism of quantum mechanics, except that the fluid function is essentially macroscopic.
It does not describe
particles, but fluid parcels which are collective degrees of freedom, while their microscopic (molecular or atomic) structure is neglected.
Nevertheless, one can still use the formal similarities  between fluid functions and wavefunctions, as will be shown below. 

\sscn{Hydrodynamic representation}{s:hdf}
Let us write the fluid wavefunction in the Madelung form 
\be\lb{e:fwf}
\Psi = \sqrt\rho \exp{(i S)}
,
\ee
where 
$S = S(\textbf{x},t)$ is a phase, which is related to fluid velocity 
\be
\textbf{u} = \cf \vena S
,
\ee
under the simplifying assumption of irrotational flow.

Substituting \eq \eqref{e:fwf} into \eq \eqref{e:o}, one obtains
\ba
&&
\partial_t\rho 
+ \vena\cdot(\rho \textbf{u})
= 0
,\lb{e:floma}\\&&
\partial_t \textbf{u}
+
\textbf{u} \cdot\vena \textbf{u}
-
\frac{1}{\rho} \vena\cdot \mathbb{T}
=0
,
\lb{e:flomo}
\ea
with stress tensor $\mathbb{T}$ of the form
\be\lb{e:stko}
\mathbb{T} 
=
-\frac{\cf}{4 \rho} \vena\rho \otimes \vena\rho 
- \tilde p \, 
\mathbb{I}
,
\ee
where $\mathbb{I}$ is the identity matrix, 
\be
\tilde p 
=
p (\rho) - \frac{1}{4} \cf \vena^2\rho 
=
- b \rho - \frac{1}{4} \cf \vena^2\rho 
\ee
is capillary pressure, 
and 
\be
p (\rho) = - b \rho
\ee
is a barotropic equation of state for the fluid 
pressure $p$.
Tensor \eqref{e:stko} belongs to the Korteweg class,
which
is used to model fluid mixtures with phase changes, capillarity effects
and diffuse interfaces \cite{ds85,amw98}.

Thus, \eq \eqref{e:o} is a concise way of writing two
hydrodynamic laws for mass and momentum conservation for a
two-phase compressible inviscid fluid with internal capillarity whose flow is
irrotational and isothermal.
This makes it useful for studies, especially
considering the large amount of analytical information about logarithmic  
wave equations which is already known.

\sscn{Phase structure and solutions}{s:phs}
It is known that Korteweg-type material can be in different phases, depending on the sign of  
nonlinear coupling  \cite{z18epl}. 
On the other hand, this coupling is also related to temperature.
Therefore, the sign of $b$
corresponds to phases occurring at the material's temperature being either above or below a certain critical value.

\textit{Cellular phase}.
For a positive coupling, 
the field-theoretical potential density is
given by
\be\lb{e:ftpot}
\lanp (|\Psi|^2) = -
b 
|\Psi|^2
\left[
\ln{(|\Psi|^2/\rho_0)} -1
\right]
+ \lanp_0
,
\ee
where $\lanp_0 = 0$.
This potential 
has an upside-down Mexican hat shape.
The local degenerate maxima of this potential are located at $|\Psi | =|\Psi_e | \equiv \sqrt{\rho_0}$.

In this case, a solitary-wave solution of \eq \eqref{e:o} 
has the form 
of a Gaussian parcel 
\ba
\Psi_g^{(+)} (\textbf{x}, t) 
&=&
\pm 
\sqrt{\rho_g^{(+)} (\textbf{x})}
\exp{\left(- i \omega_g^{(+)} t
\right)} 
,\lb{e:solg}\\
\rho_g^{(+)} (\textbf{x})
&=&
\tilde\rho
\exp{\left[- \frac{(\textbf{x}- \textbf{x}_0)^2}{\ell^2} \right]}
,\lb{e:gauss}\\
\omega_g^{(+)}
&=&
b 
\left[
\dm - 
\ln{
 \left(
        \frac{\tilde\rho}{\rho_0}
 \right)}
\right]
,
\ea
where $\dm$ is the number of spatial dimensions,
and
\be\lb{e:ell}
\tilde\rho = M/\tilde\vol,
\
\tilde\vol= 
\pi^{\dm/2} \ell^\dm
,
\
\ell = \sqrt{|\cf/(2 b)|}
\ee
are the 
density peak value, effective volume and Gaussian width, respectively.

The solution $\Psi_g^{(+)}$ corresponds to the state with the lowest frequency eigenvalue, $\omega^{(+)}$, which makes it
analogous to the ground state in wave mechanics. 
It has been extensively studied since Rosen's work \cite{ros68}.


Thus, if nonlinear coupling is positive, then our material tends to
fragment into clusters of density inhomogeneities 
with a Gaussian profile each, referred to here as cells.
In the multi-solution picture, this results in a phase which is a composition of Gaussian-like cells.
If this phase occurs in the melt of a metal, then during the solidification process the central regions of these cells are likely to become
nucleation centers; resulting in a polycrystalline structure.

\textit{Foam phase}.
If the nonlinear coupling
is negative, then
the potential  \eqref{e:ftpot}
has a habitual Mexican-hat shape,
assuming $\lanp_0= - b \rho_0 = |b| \rho_0$.
Its local maximum is located at $|\Psi | = 0$
and local degenerate minima are located at $|\Psi | =|\Psi_e | $. 
These minima correspond to the state with the lowest eigenvalue of frequency $\omega_{g}^{(-)} = 0$.
This indicates the presence of multiple topological sectors in the model,
as well as topologically nontrivial solitons which interpolate between these
minima.

To the best of our knowledge, no topological solutions have been found 
in an analytical form,
but 
their existence
can be revealed by numerical studies.
In $\dm$-dimensional space,
in Cartesian coordinates, this solution is searched in the form of a product of $\dm$ 1D kink solitons 
\be
\Psi_s^{(-)} (\textbf{x}, t) = \prod\limits_{j=1}^\dm \psi_j (x_j, t)
,
\ee
and
each of these 1D solitons saturates the Bogomolny-Prasad-Sommerfield (BPS) bound
and has a nonzero topological charge,
$ 
Q = \varrho_0^{-1/2}
\left[
\psi_j(x_j \to +\infty) - \psi(x_j \to -\infty)
\right]
,\ 
\ j=1,..,\dm
$, 
where $\varrho_0 = \rho_0^{1/\dm}$.
The plots of numerical solutions and corresponding densities can be found  
in \Fig 2 of \Ref\cite{z18epl}.

Depending on the value of $Q$, all non-singular finite-energy solutions 
can be cast into four
topological sectors
\cite{z11ap}.
For two of these sectors, the topological charge is not zero, ensuring the stability
of the corresponding solitons.
Thus,
despite the kink solution being not the lowest frequency one,
its stability against decay into a ground state $\Psi_e $ is enhanced by a nonzero topological charge.

One can check that the mass density of the soliton, given by $\bigl| \Psi_s^{(-)}\bigr|^2 $, 
grows from its center of mass outwards, 
cf. \Fig 2 of \Ref\cite{z18epl}.
Therefore, in three dimensions
the
solution $\Psi_s^{(-)} $ can be viewed
as describing a bubble with a characteristic size $\ell$.
In a single-solution setup, it would fill the entire space, 
but in 
a multi-soliton picture of real material,
kinks and antikinks would alternate and match 
at distances of an order $\ell$.

In other words, in this phase
our material forms foam-like structures,
which facilitates the release of previously dissolved gas.
Such a process can result either in boiling during the liquid-gas transition
(if the temperature of the foam phase is larger
than the temperature of the cellular phase),
or in the formation of cavities in the bulk of the material, 
such as the pores and blowholes
caused by decreased solubility and production of pertinent gases during solidification
(if the temperature of the foam phase is smaller
than the temperature of the cellular phase).

To summarize, depending on whether the temperature of the foam phase is larger or smaller
than that of the cellular phase,
the two-phase structure of our model can be used for describing both liquid-gas 
and liquid-solid transitions with formation of bubbles or cavities.

\begin{figure}[t]
\centering
\subfloat[(a) steel S235
]{
  \includegraphics[width=\sc\columnwidth]{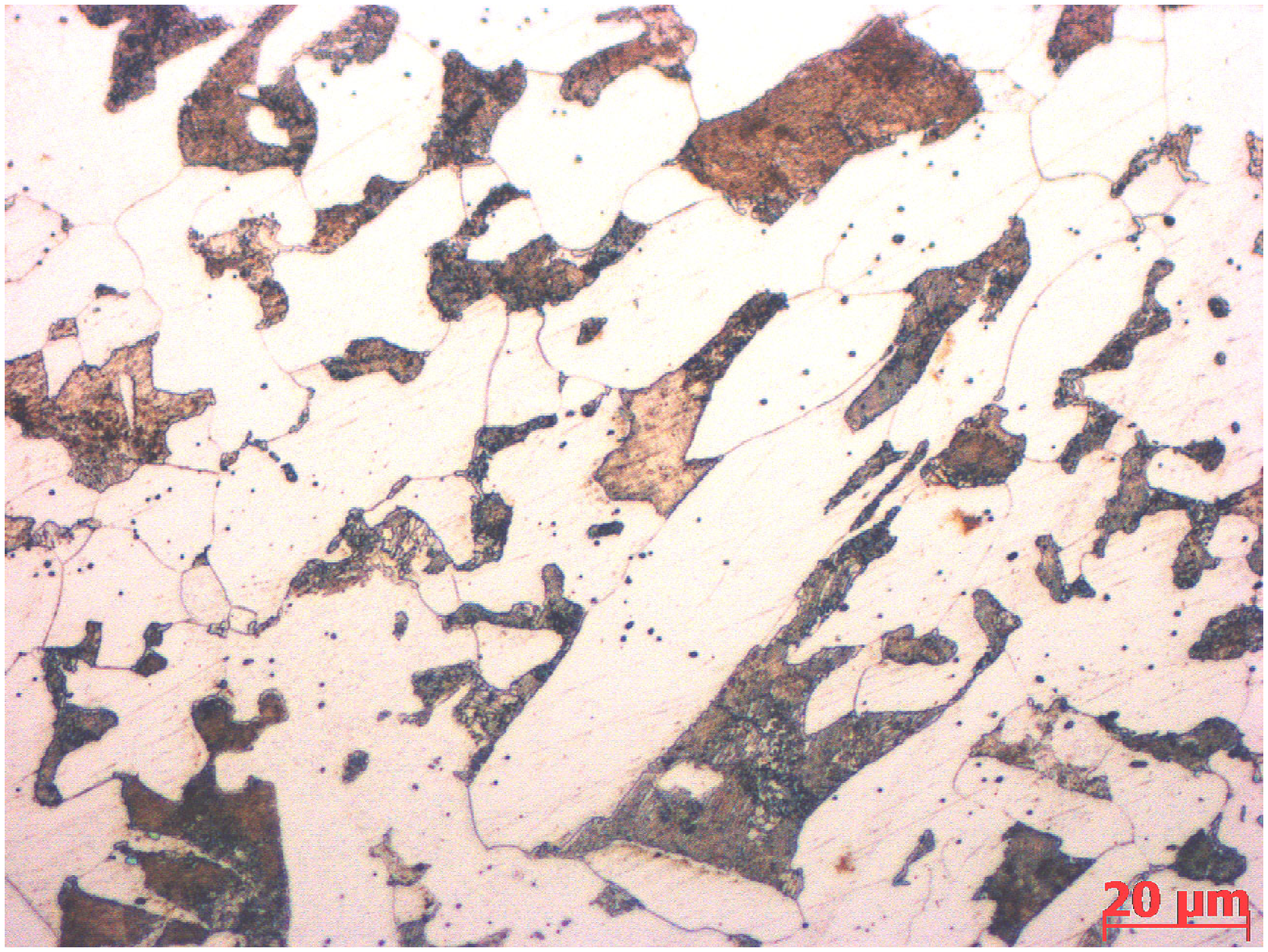}
}
\subfloat[(b) copper alloy C-Cu
]{
  \includegraphics[width=\sc\columnwidth]{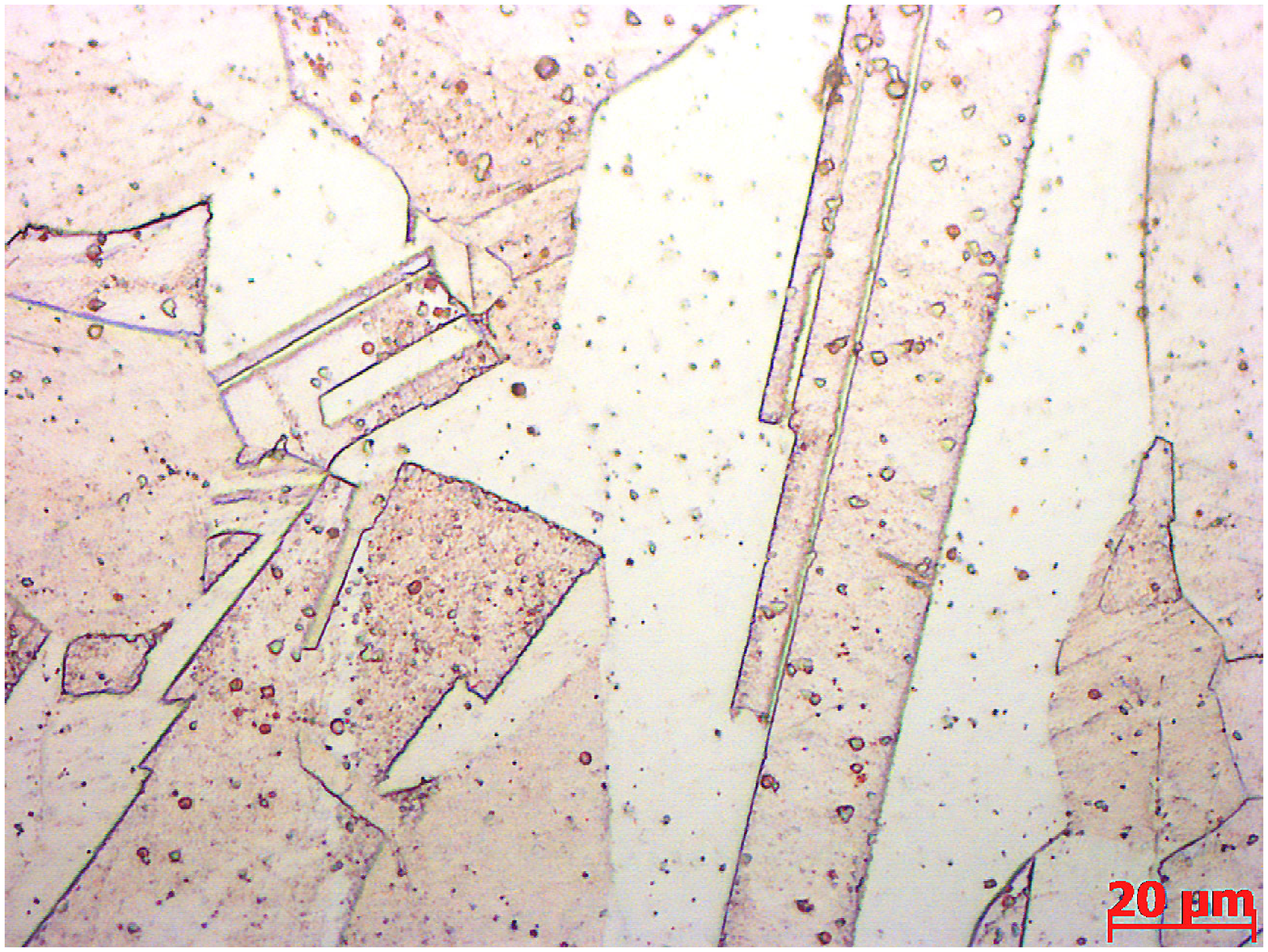}
}
\hspace{0mm}
\subfloat[(c) stainless steel X10CrNiTi18-10
]{
  \includegraphics[width=\sc\columnwidth]{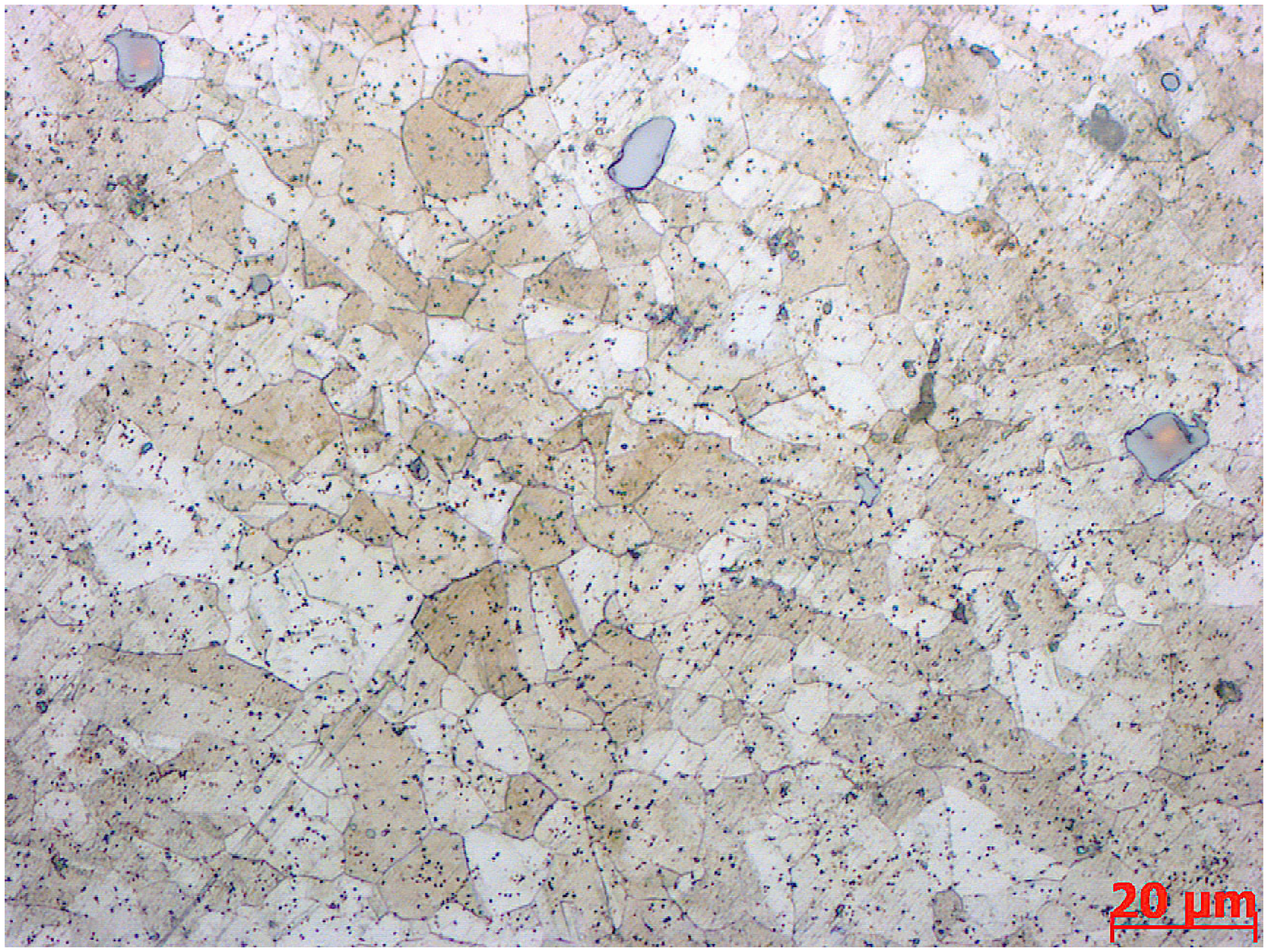}
}
\subfloat[(d) aluminium alloy 5083
]{
  \includegraphics[width=\sc\columnwidth]{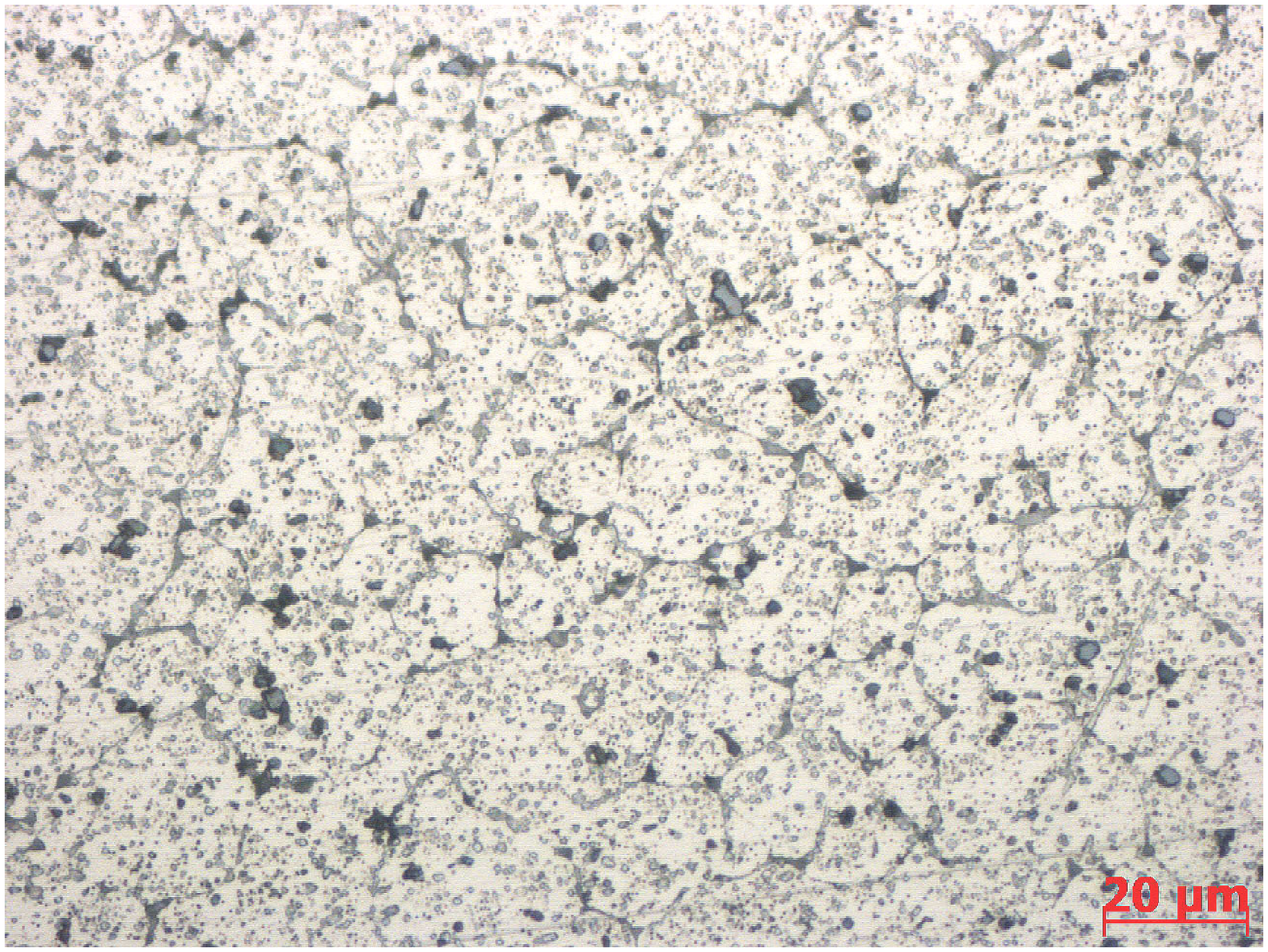}
}
\caption{Microstructure of the studied materials.
}
\label{f:mcstr}
\end{figure}

\scn{Theoretical predictions}{s:thp}
When applied to liquid-solid phase transitions in polycrystalline metals,
the logarithmic wave-mechanical model answers a number of the questions
raised in the Introduction, such as: 
is there any large-scale pattern in those metals' structure,
and how one can derive or estimate hardness and density distributions.
We shall deal with the application of our model to these problems, to more closely examine how it resolves them in Sections \ref{s:ls} and \ref{s:hdn}, respectively.

\sscn{Large-scale structure of metals}{s:ls}
Let us assume that the melt can be described as a logarithmic Korteweg-type material, at least in 
a leading-order approximation.
Then according to our model,
its large-scale density profile 
has a pattern consisting of repeating solitary-wave (gaussons) or alternating (kinks-antikinks) solitonic solutions, depending on the phase
our material is in.
In other words, density inhomogeneities, which are not 
directly related
to molecular bonding but are a collective nonlinear wave-mechanical phenomenon,
emerge in a melt at a larger length scale.

In the case of metals undergoing a solidification process,
those points
where the density profile reaches its extrema, 
are more likely
to become the centers of grains' nucleation.
Therefore, an average grain size is not random,
instead its value must be close to the width 
$\ell$ given by equation 
\eqref{e:ell}.

\sscn{Microhardness}{s:hdn}
Apart from the non-random distribution of grains
at a large scale (about ten times the size of a grain,  and above),
our model predicts a Gaussian-like distribution of matter density inside a grain,
\textit{i.e.}, from
its center to the boundary.
This has implications for the distribution of microhardness inside a grain,
for the following reasons.

The non-uniform distribution of mechanical properties and density inside a grain is known to occur 
as a result of the following effects:
segregation of chemical elements and repulsion  of impurities from the
center of a grain to its
boundary during the crystallization process,
the formation of stresses, as well as defect clusters of the crystal structure (dislocations),
at the grain's boundary \cite{sal64,nz66,fpk12,ktz06}.
These processes start to occur in the melt, alongside the formation of crystals' seeds. 
Strength distribution depends on the stage the crystallization process is in,
as well as on the quantity and size of seeds and liquid metal zones between them \cite{fle74}.
The crystals thus become spatial regions with larger hardness.

A qualitative correspondence exists between non-uniform distributions of 
strength and local density inside a grain.
Variations of local density across a grain,
and 
the formation of regions of higher and lower density
are caused by the accumulation of crystal grate defects, such as  
dislocations and vacancies. 
The higher probability of defects being located towards the grain's boundaries
leads, on average, to lower values of density at the boundaries \cite{ggm74,kv87}. 

The redistribution of density inside the melt region of a size comparable to the size of 
the future grain 
occurs during the melt's cooling and crystallization.
Metals have a lower density in a liquid state than in a solid one,
cf.
steel: 6500-6800 versus 7700-8000 kg/m$^3$, 
copper: 8217 versus 8930 kg/m$^3$, 
and
aluminium: 2382 versus 2700 kg/m$^3$. 

As the melt cools, 
the growth of crystals from crystallization centers increases
the volume of denser matter, until the boundaries of different grains touch each other.
In the crystal which is thus forming,
the redistribution of density occurs from its center to its boundary (with a
minimum at the boundary), due to the above-mentioned effects as a result of defects.
This is a continuous process
until the temperature reaches about half of the alloy's 
melting temperature.
In a completely solidified grain,
hardness and density both follow a normal distribution, but in an anti-correlated way:
hardness has a maximum value at the grain's boundary,
whereas density achieves its maximum in the center.
The plastic deformation of a grain might increase the amplitude of this distribution, but it
does not qualitatively
change its main feature. 

Therefore, our model predicts the periodic distribution of hardness across the specimen,
with the average period being equal to the average size of a grain;
and both of these values being equal to the characteristic length scale $\ell$. 

In the next section, we compare these predictions with experimental data.

\scn{Model-experiment comparison}{s:ex}
%
%
To confront our model predictions with experiment,
we study a number of the most common polycrystalline metallic materials:
structural steel S235/A570, copper C-Cu/C14200, stainless steel X10CrNiTi18-10/AISI 321, and aluminium-magnesium alloy 5083/5056.

The microstructure of steel S235/A570 consists of the ferrite (light regions in \Fig\ref{f:mcstr}a)
and pearlite (dark regions in \Fig\ref{f:mcstr}a) phases;  
their ratio is about $75/25$, respectively.
Ferrite grains mainly have  an equiaxial structure, with the exception of a few stringer grains of an elongated shape.
They have no particular spatial orientation, and
their size is about $20$ $\mu$m.
Sizes of pearlitic microconstituent grains for this steel vary between $5$ and $45$ $\mu$m.

The microstructure of the copper alloy specimen consists of approximately equiaxed grains 
of a size about $100$ $\mu$m, see \Fig\ref{f:mcstr}b.
Similarly to the previous material,
a few stringer grains of an elongated shape are present.
Also, deformation twins can be found inside the grains.

The microstructure of the stainless steel specimen is austenitic,
and consists of polyhedral grains of a size of about $100$ $\mu$m on average, \Fig\ref{f:mcstr}c.
As in the copper specimen, one can also encounter deformation twins in this steel.

Finally,
the microstructure of the aluminium alloy specimen consists of approximately equiaxed grains 
of a size about $30$ $\mu$m, see \Fig\ref{f:mcstr}d.
A small quantity of secondary ferric and silicon phases
can be found both inside the grains and on their boundaries. 

Based on this data and using \eq\eqref{e:ell},
one can deduce a bound for the parameters of the models for each of the specimens.
We obtain that
$|\cf/b| \approx 0.8 \times 10^{-9}$ m$^2$ for ferrite steel,
$|\cf/b| \approx 2 \times 10^{-8}$ m$^2$ for copper alloy and stainless steel,
and
$|\cf/b| \approx 1.8 \times 10^{-9}$ m$^2$ for aluminium alloy.

\begin{figure}[t]
\centering
\subfloat[]{
  \includegraphics[width=\lsc\columnwidth]{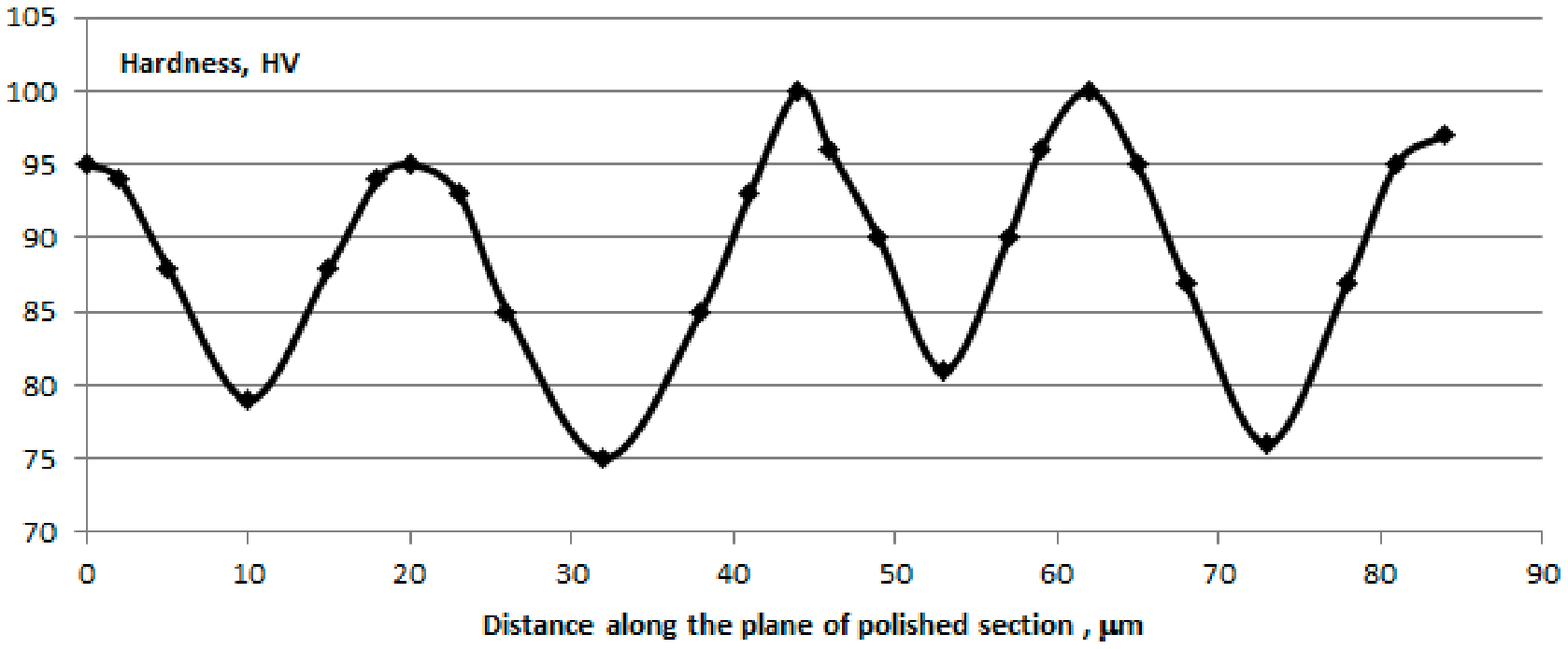}
}
\subfloat[]{
  \includegraphics[width=\rsc\columnwidth]{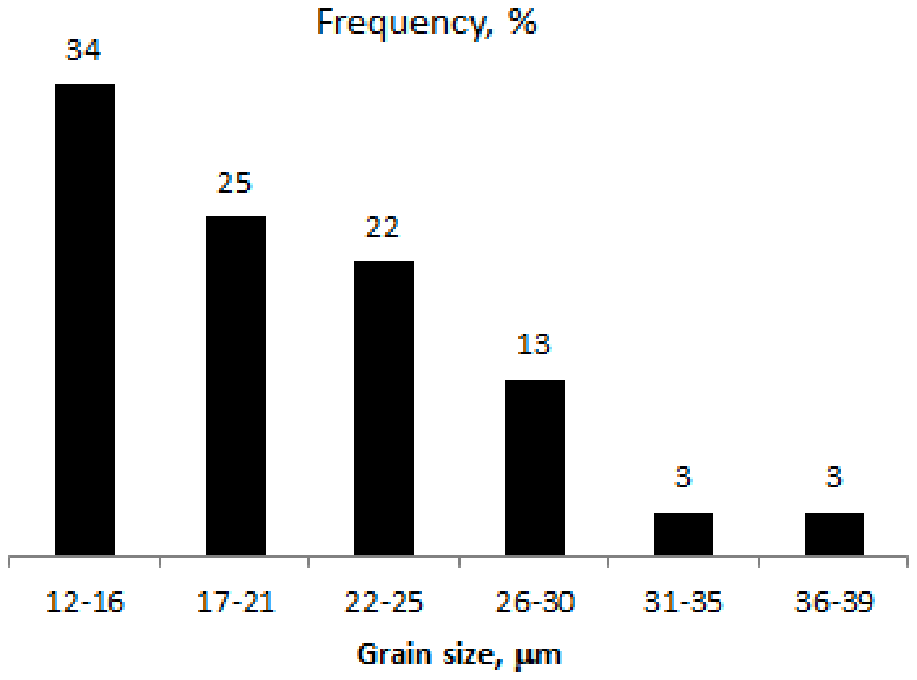}
}
\hspace{0mm}
\subfloat[]{
  \includegraphics[width=\lsc\columnwidth]{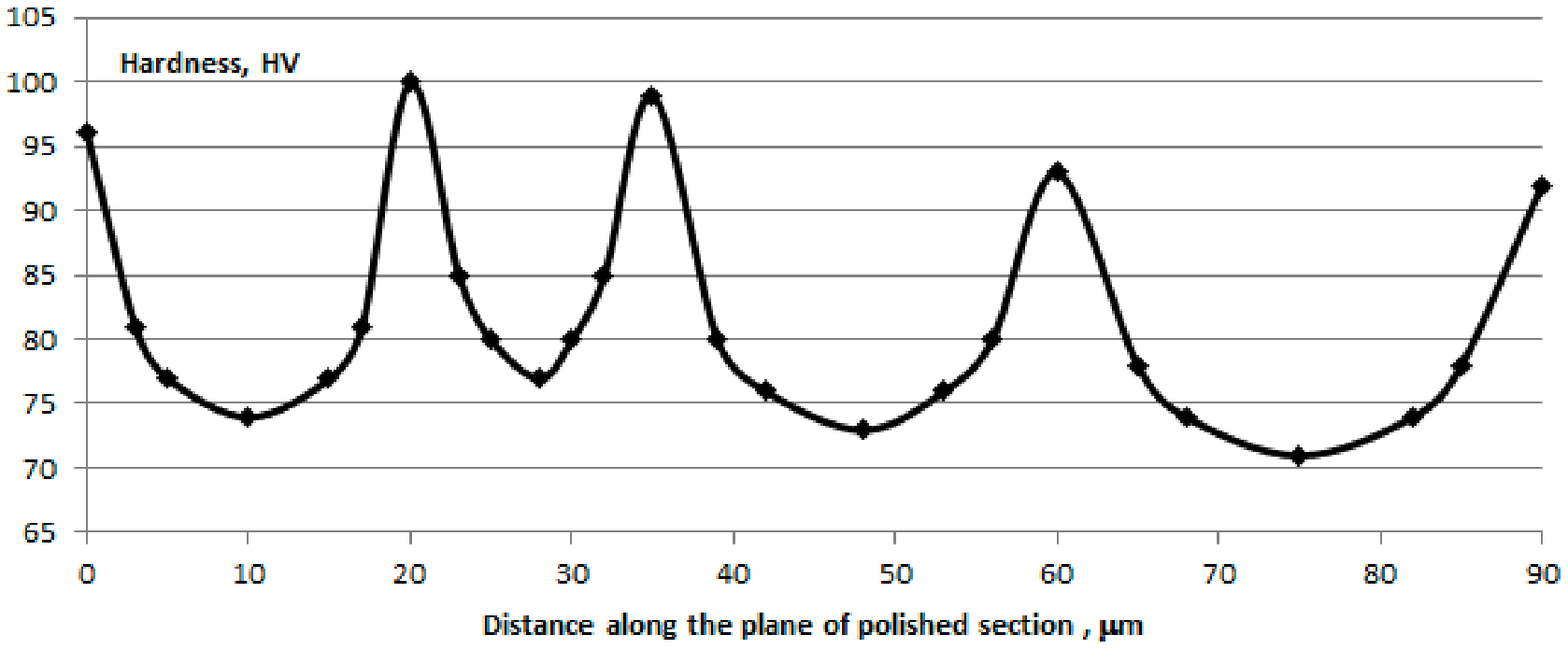}
}
\subfloat[]{
  \includegraphics[width=\rsc\columnwidth]{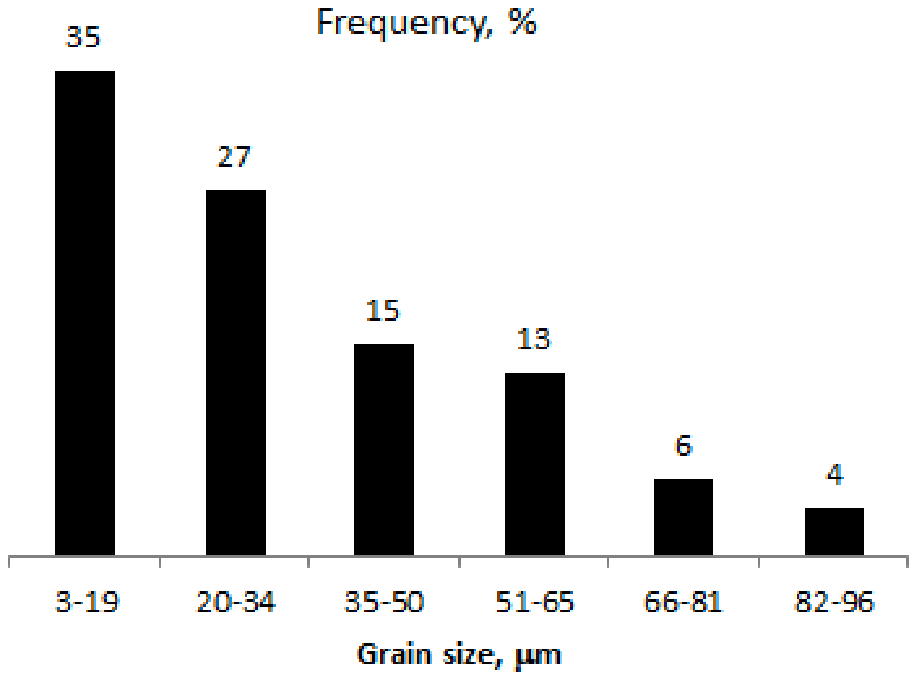}
}
\hspace{0mm}
\subfloat[]{
  \includegraphics[width=\lsc\columnwidth]{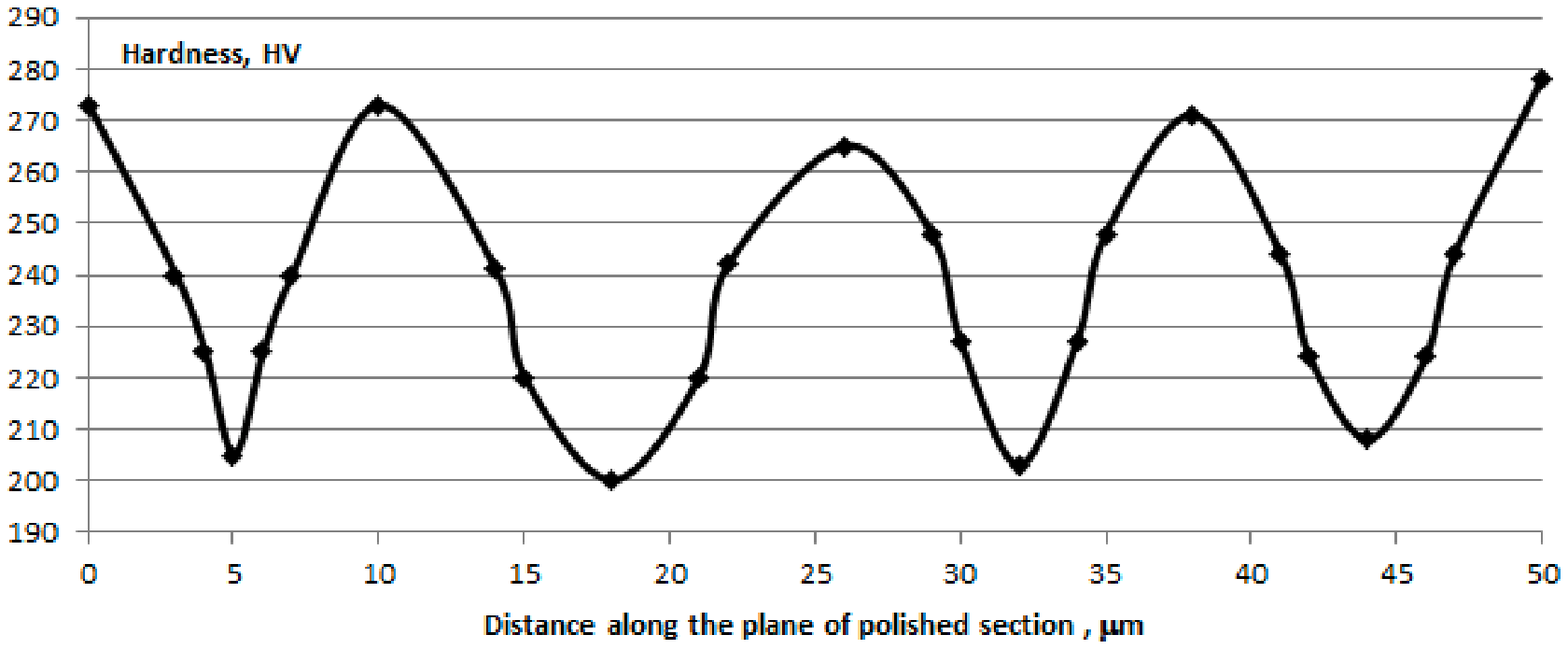}
}
\subfloat[]{
  \includegraphics[width=\rsc\columnwidth]{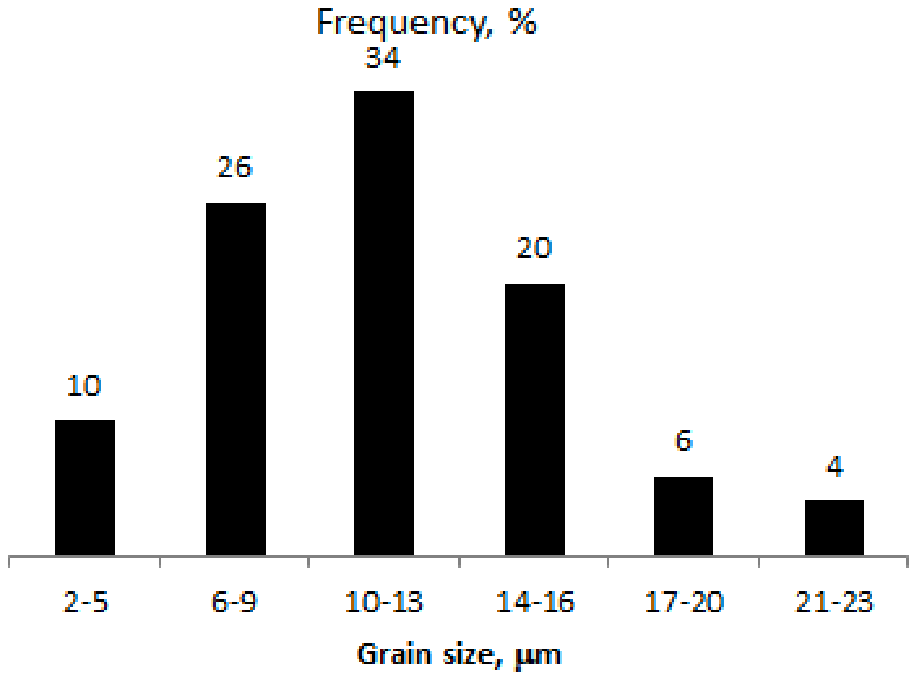}
}
\hspace{0mm}
\subfloat[]{
  \includegraphics[width=\lsc\columnwidth]{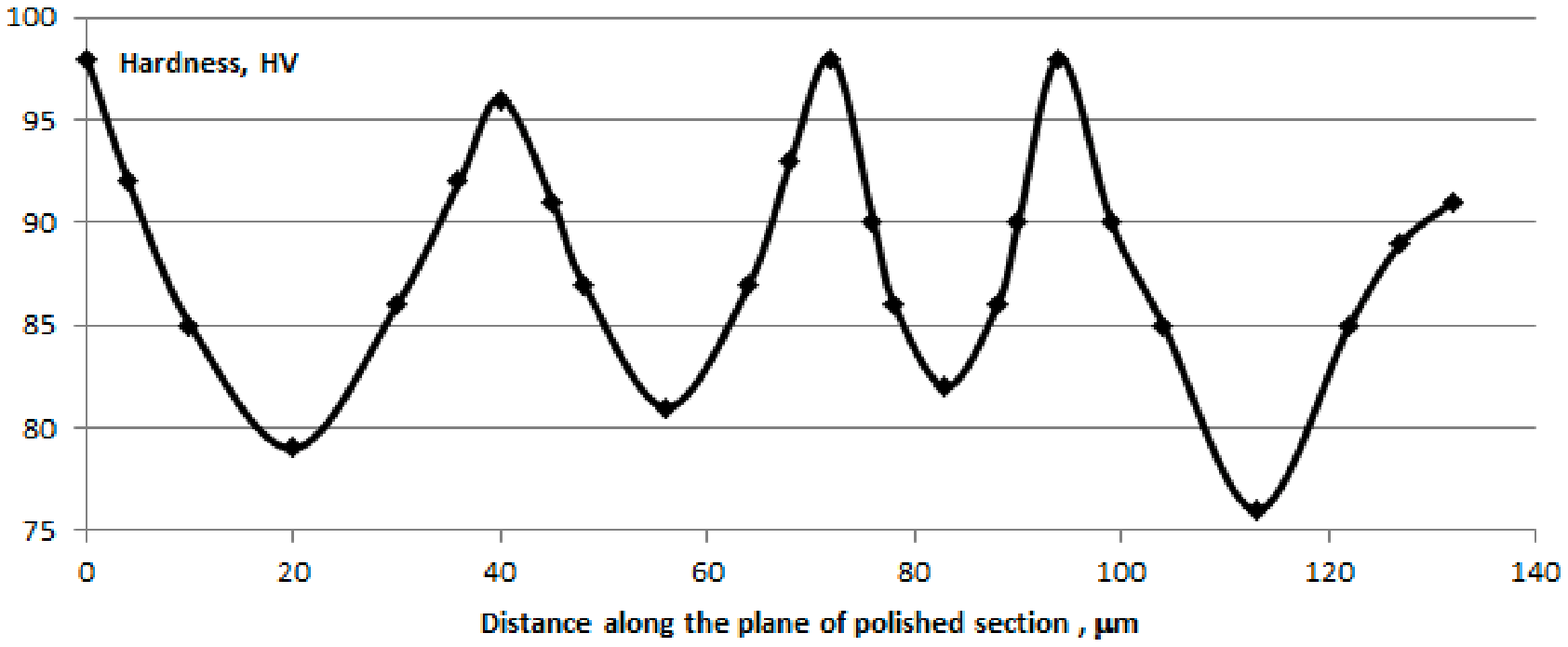}
}
\subfloat[]{
  \includegraphics[width=\rsc\columnwidth]{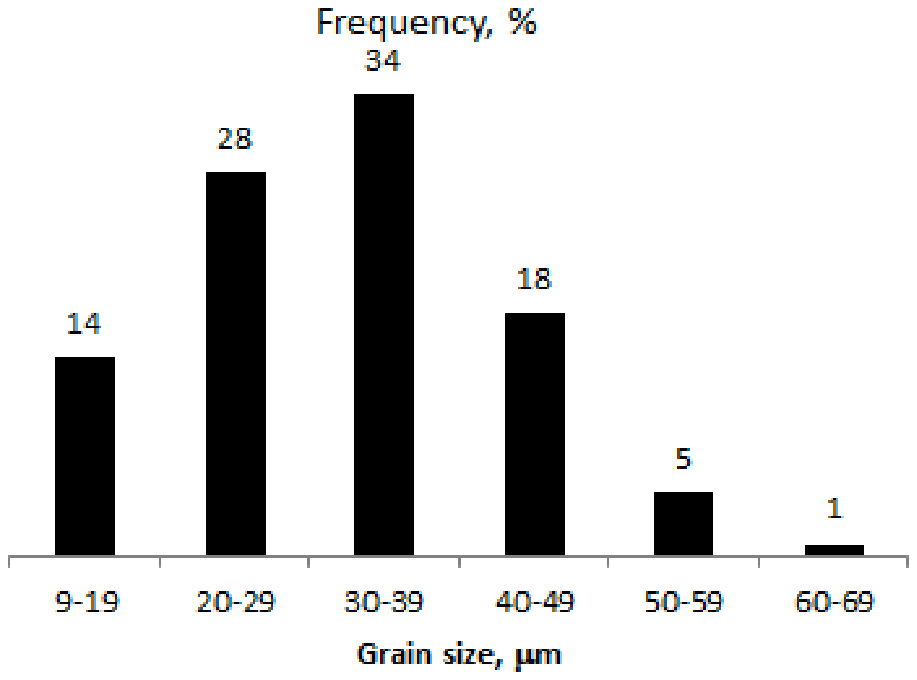}
}
\caption{Hardness profiles (left column) and grain size distributions (right column) in, 
from top to bottom: ferrite S235, copper alloy C-Cu, austenite X10CrNiTi18-10, and aluminium alloy 5083.}
\label{f:size}
\end{figure}

Furthermore,
microhardness measurements 
were performed for the above-mentioned materials, using the Vickers hardness method.
The load during the measurements was chosen according to the size of 
an indentation
in a specimen.
The test force was chosen to be either 0.025 or 0.05 N, depending 
on the grain's size,
according to ISO 6507-1:2005(E) specifications.
The points of the load's application were in the following areas of the grain:
the
grain's center, half-way to the grain's boundary, near the boundary, and on the boundary.
Measurements were performed with a statistical dataset of 40-90 points per sample.
The subsequent statistical analysis was performed 
at a confidence level of 95 $\%$, see Table \ref{t:hard}.
The profiles of microhardness measured across each specimen for the chosen materials are presented 
in
\Fig\ref{f:size}.

\bw\bc
\begin{table}\centering
\caption{Microhardness distribution inside a grain.}\label{t:hard}
\begin{tabular}{|c|cccc|}
\hline
\multicolumn{1}{|c|}{~}
&
\multicolumn{4}{c|}{\textbf{Microhardness} (HV)}\\
\cline{2-5} 
\textbf{Alloy}&~~~~~~at the~~~~~~&~~~halfway to ~~~&~~near the~~&~~~~~~at the~~~~~~\\
&~center~&boundary&~boundary~&~boundary~\\
[1mm]
\hline
Ferrite S235&$77\pm 3$&$87\pm 3$&$94\pm 3$&$98\pm 4$\\
[1mm]
Copper C-Cu&$74\pm 3$&$78\pm 3$&$81\pm 3$&$96\pm 4$\\
[1mm]
~Austenite X10CrNiTi18-10 ~&$205\pm 5$&$224\pm 5$&$244\pm 6$&$269\pm 9$\\
[1mm]
Aluminium 5083&$79\pm 3$&$87\pm 2$&$92\pm 2$&$98\pm 3$\\
[1mm]
\hline
\end{tabular}
\end{table}

\ec\ew

Let us compare the experimental results with our
theoretical predictions.
Because  hardness is related to the density of a material,
theory predicts the occurrence of a periodic structure at length scales
which are larger, by an order of magnitude,
than the size of an average grain.
The experimental data confirms this conjecture,
see the left-hand side plots in \Fig\ref{f:size}.
Statistical analysis confirms that the average size of the grains is
not random: its distribution is far from being uniform, but always
has a clear global maximum, regardless of the material,
cf. the right-hand side plots in \Fig\ref{f:size}.

Furthermore,
if the distribution of hardness inside each grain is anti-correlated with the distribution
of density, the
data confirms that the latter must have a Gaussian-like form, as predicted by the model.
The measured average sizes of the grains thus provide an experimental bound
for
the Gaussian widths of the soliton-type solutions $\ell$ for each material studied.

%

\scn{Conclusion}{s:dis}
In this \art,
we considered the so-called logarithmic Korteweg-type materials,
which are hydrodynamic models using
wave equations with logarithmic nonlinearity.
Their corresponding solutions have either trivial 
and non-trivial topology - when coupling is, respectively, positive or negative.

These models were applied to various materials,
such as polycrystalline alloys, which undergo liquid-solid or liquid-gas phase transitions. 
Such materials have a two-phase structure, where each phase is labeled by a sign of the
coupling of logarithmic nonlinearity and described by above-mentioned solutions.

From a physical point of view, 
such solutions 
describe the density inhomogeneities, which can manifest themselves in the form of either bubbles (cavities) or cells 
in the vicinity of the liquid-gas (liquid-solid) phase transition. 
In materials undergoing a solidification process,
these inhomogeneities become centers of nucleation of grains occurring in the liquid phase
in the vicinity of a melting point.
One of the theory's predictions is a large-scale periodical Gaussian-like pattern,
which occurs instead of expected randomness, if based on conventional statistical arguments.
Due to  additional processes and effects occurring during solidification, this periodicity can distorted,
but not entirely destroyed.

The single-peaked (predominantly Gaussian) behaviour of the distribution of grain sizes 
was established phenomenologically a long time ago \cite{sal58,gor78},
but its statistical-mechanical theoretical explanation was hitherto unknown, 
to the best of our knowledge.
In this paper, the occurrence of Gaussian-like repeating patterns is not only 
explained on hydrodynamic and statistical-mechanical grounds, but also used
for describing density and hardness of polycrystalline metals.
This illustrates the importance of taking into account properties of non-solid (liquid and gas)
phases of a solidifying material 
when trying
to model its solid phase's mechanical properties.

In previous works, natural silicate materials, such as magmas in volcanic conduits, were considered \cite{dmf03,z18epl},
and the (approximately) periodical flows and structures were established.
In this paper, we reported  experimental studies of the structural (quasi)periodicity, microhardness, density and size of grains in
structural steel S235/A570, copper C-Cu/C14200, stainless steel X10CrNiTi18-10/AISI 321, and aluminium-magnesium alloy 5083/5056.

To conclude,
the wave-mechanical logarithmic model explains spatial variations in density and hardness of polycrystalline metals
originating from their liquid phase during the solidification process.
At a microstructural level, domains of spatial density variations form, 
which have Gaussian profiles of values,
and the peaks of their distribution functions strongly affect the polycrystalline structure of the metal grains.
The obtained results, along with their comparison with experiment,  indicate an applicability of the model 
of a Korteweg type to a wide range of polycrystalline alloys.
This allows us to provide analytical estimates for some of their mechanical characteristics,
which are most important for applications.

\begin{acknowledgments}
K.G.Z. is grateful to participants of the
XXVI International Conference on Integrable Systems and Quantum Symmetries (ISQS-26)
for stimulating
discussions and remarks \cite{kkd19}.
This research is supported 
by the Department of Higher Education and Training of South Africa
and
in part by the National Research Foundation of South Africa (Grants Nos. 95965, 132202 and 131604).
Proofreading of the manuscript by P. Stannard is greatly appreciated.\\
\end{acknowledgments}

\noindent
\textbf{\small Compliance with Ethical Standards}\\~\\
\noindent
\textbf{\small Availability of data and materials}
{\small The data may be available from the first author upon request.}\\
\textbf{\small Conflict of interest} {\small The authors declare no conflict of interest}.\\


\end{document}